\documentstyle[12pt,aasms4]{article}

\lefthead{}
\righthead{}

\slugcomment{In preparation for {\it The Astronomical Journal}}

\begin{document}

\title{The Metallicity and Reddening of Stars in the Inner Galactic Bulge}

\author{Jay A. Frogel\altaffilmark{1}, Glenn P. Tiede\altaffilmark{2}, \& Leslie E. Kuchinski\altaffilmark{3}}
\affil{Department of Astronomy, The Ohio State
University, 174 W. 18th Avenue, Columbus, Ohio  43210}
\authoremail{\{frogel@astronomy.ohio-state.edu, tiede@noao.edu, leslie@ipac.caltech.edu}

\altaffiltext{1}{Visiting Research Associate, The Observatories of the 
Carnegie Institution of Washington.}
\altaffiltext{2}{NOAO, 950 N. Cherry Ave., Tucson, AZ  85719-4933}
\altaffiltext{3}{Current Address: IPAC, California Institute of Technology, Pasadena, CA 91125}

\begin{abstract}

We present a preliminary analysis of $K$, $J-K$ color magnitude
diagrams (CMDs) for 7 different positions on or close to the minor axis
of the Milky Way at Galactic latitudes between $+0.1^\circ$ and
$-2.8^\circ$.  From the slopes of the (linear) giant branches in these
CMDs we derive a dependence of $\langle$[Fe/H]$\rangle$ on latitude for
$b$ between $-0.8^\circ$ and $-2.8^\circ$ of $-0.085 \pm 0.033$
dex/degree.  When combined with the data from Tiede {\it et al.} we
find for $-0.8^\circ \leq b \leq -10.3^\circ$ the slope in
$\langle$[Fe/H]$\rangle$ is $-0.064 \pm 0.012$ dex/degree.  An
extrapolation to the Galactic Center predicts [Fe/H] $= +0.034 \pm
0.053$ dex.

We also derive $average$ values for the extinction in the $K$ band
($A_K$) of between 2.15 and 0.27 for the inner bulge fields studied,
corresponding to average values of E($J-K$) of between 3.46 and 0.44.
There is a well defined linear relation between the average extinction
for a field and the star-to-star scatter in the extinction for the
stars within each field.  The equation of this line is
$\sigma(A_K)=0.056(\pm 0.005) \langle A_K \rangle + 0.043(\pm 0.005)$.
This result suggests that the typical apparent angular scale size for
an absorbing cloud is small compared with the field size ($90\arcsec$
on a side).

Finally, from an examination of the luminosity function of bright
giants in each field we conclude that the
young component of the stellar 
population observed near the Galactic center declines in density much
more quickly than the overall bulge population and is undetectable beyond
$1^\circ$ from the Galactic center.

\end{abstract}


\section{Introduction}
 
In a seminal paper, \cite{whi78} demonstrated that the integrated
optical light from Baade's Window in the Galactic bulge ($b =
-4^\circ$) closely resembles the light from the bulges of other spirals
and from moderate luminosity E and S0 galaxies.  Contemporaneous with
Whitford's paper the near-IR (1.2 to 2.2 $\mu$m) light from these
galaxies was shown to be dominated by giant stars (\cite{fro78}).  A few years later, Blanco and his collaborators (Blanco
{\it et al.} 1984) established that Baade's Window contains a high
percentage of middle and late M giants compared to the mix of stars in
other parts of the Galaxy.  Subsequent optical and IR studies of these
bulge giants (\cite{fro88}; Frogel \& Whitford 1987; \cite{fro90};
\cite{ric83}; Rich 1988; Terndrup et al. 1990, 1991) led to the
conclusion that their photometric and spectroscopic properties
corresponded to what was expected for the stars inferred to exist in
early-type galaxies based on the integrated light observations of the
latter.  At the same time,  these optical and IR studies revealed that
the observed luminosities, colors, and spectral characteristics of the
bulge stars were significantly differerent from those of other known M
giants.

Nearly all of our present knowledge of stars in the Galactic bulge has
been derived from observations of stars in fields with $\vert b \vert
\geq 3^\circ$, i.e. greater than about one scale length.  For example,
Tiede {\it et al.} (1995) compiled mean metallicity values for outer
bulge fields in order to determine the best estimate for the gradient in
metallicity over this region.  Minniti {\it et al.} (1995) discussed the evidence for and meaning of a metallicity gradient in the Galactic bulge but again based on
observations of fields exterior to Baade's Window.  For $\vert b \vert
\leq 3^\circ$, on the other hand, reddening and extinction increase
strongly so that optical observations become difficult to impossible.
Near-IR observations then become the only way that these inner bulge
stars can be studied effectively, either photometrically or
spectroscopically (e.g., \cite{gla87}; \cite{leb87}; \cite{cat89}).

Most of our previous near-IR observations of bulge giants have been in
the same fields that have been studied optically.  This has allowed us
to develop calibrated techniques for estimating key physical parameters
for these stars such as reddening,  metallicity, and luminosity based
on both photometry and low to medium resolution spectroscopy, both in
the near-IR.  For example, Tiede {\it et al.} (1995) used the
relationship between [Fe/H] and the slope of the upper giant branch
(GB) in a $K, J-K$ color-magnitude diagram (CMD) derived from
observations of globular clusters by \cite{kuc95} to estimate the mean
metallicity of several fields along the minor axis of the bulge.  They
found a gradient in [Fe/H] of $-0.060 \pm0.033$ dex/degree for the
region $-12^\circ \leq b \leq -3^\circ$.

If we are to achieve a better understanding of the formation,
evolution, and chemical enrichment history of the bulge we must observe
stars much closer in to the center.  As noted above, these observations
are best carried out in the near-IR.  The analysis techniques developed
with stars in globular clusters and the outer bulge can then be applied
to these data.  We have already obtained most of the needed {\it JHK}
images of selected inner bulge fields.  From these data we can
calculate stellar luminosities and reddenings, derive luminosity
functions, and identify candidate stars for near-IR spectroscopy.  We
are now in the process of obtaining near-IR spectra and establishing a
metallicity calibration for these data from similar observations of
giants in metal-rich globular clusters.

The purpose of this paper is to present a preliminary analysis of
near-IR CMDs for 11 fields interior to $-4^\circ$ and as close as
$0.2^\circ$ to the Galactic Center itself.  Seven of these fields are
on the minor axis; 5 are at a latitude of $-1.3^\circ$ parallel to the
major axis.  From these CMDs we first estimate (sections 3.1 and 3.2)
the reddening for each field with a technique similar to that employed
by \cite{nar96}.  We then estimate the mean metallicity for each field
(section 3.3) using the slope of the giant branch method.  This allows
us to improve the reliability of our earlier value for the metallicity
gradient in the Galactic bulge (sections 3.4 and 3.5) since our
previous data set included only one inner bulge field.  While
metallicities derived with this technique may not be as accurate as
those that can be obtained from spectroscopy, given the impossibility
of obtaining ``definitive'' metallicities from high resolution optical
spectroscopy and the relative newness of techniques based on near-IR
spectroscopy, application of several different, independent methods to
the problem is valuable.  Also, we emphasize that although we will
refer to metallicity as [Fe/H] in this paper, and we have based our
metallicity scale on observations of globular clusters, most of which
have had a true Fe abundance determined, the slope of the giant branch
could be affected by non-solar abundance ratios with respect to Fe (cf.
\cite{mcw94}) in the bulge.  Finally, from a ``quick look'' at the
luminosity functions of the bright giants in each field we are able to
address the issue of whether or not there has been a significant amount
of continuing star formation in the bulge (section 3.6).  In subsequent
papers, we will give a detailed analysis of the $JHK$ colors and
luminosities of the giants in these fields and derive independent
[Fe/H] estimates from near-IR spectroscopy of the brighter giants in
each of the fields.

\section{Observations and Data Reduction}

All of our data were obtained on the 2.5m duPont telescope at Las
Campanas Observatory during 1992 and 1993 with IRCAM (\cite{per92}).
The detector was a $256 \times 256$ HgCdTe NICMOS 3 array with a plate
scale of 0.348 arcsec pixel$^{-1}$.  The log of the observations is
given in Table~\ref{obsdat}.  Figure~\ref{fields} displays the location
of the fields.  We have included the location of Baade's Window in this
figure to underscore the fact the new fields provide a good sampling of
the entire bulge minor axis interior to  Baade's Window, a region of
the Galaxy that, except for the Galactic Center itself, is almost
completely unexplored.  For the 3 fields closest to the Galactic center
(the ``c'' fields) we chose regions of relatively uniform extinction
based on visual inspection of near-IR images of the immediate vicinity
of the Galactic Center (Glass {\it et al.} 1987), originally presented
in \cite{cat85}.  We note, though, that compared to fields at higher
latitudes away from the center, even these fields have very patchy
extinction.  In all of our logs and data reduction procedures these
fields were always referred to as the ``Catchpole fields'', hence the
``c'' designation.  The data analyzed in \cite{tie95} were obtained as
part of this program and the observations were made on the same
nights.  Data reduction proceeded as described in Tiede {\it et al.}
except for the differences we will now note.

Because we wanted to well sample the upper parts of the GB in each of
these fields and because the field of view of
IRCAM is small relative to the areas of the higher latitude fields 
studied by Frogel {\it et al.} (1990), we imaged one part of each field deeply
and then took shorter exposures in a grid pattern, usually $70\arcsec$
on a side.  This strategy resulted in the a--d designations in 
Table~\ref{obsdat}.  The second and third columns of Table~\ref{obsdat}
give the coordinates of the nominal field centers together with the offsets
with respect to these centers of the adjacent fields.  For various
reasons, the data from the adjacent fields were not always usable, hence
the incomplete coverage in Table~\ref{obsdat}.  Usually, there was an 
overlap region between 10 and $20\arcsec$ for adjacent fields.  We
caution observers who might use the coordinates in Table~\ref{obsdat}
that their accuracy is probably not better than $\pm10\arcsec$.

The $5^{\rm th}$ and $6^{\rm th}$ columns of Table~\ref{obsdat} give the
number of images taken and the exposure times.  For all pointings, we
took 5 images with centers displaced by a few seconds of arc from each
other.  Short, one second, exposures were always taken to minimize saturation
effects on the brightest stars.  This series was then immediately followed
by a series of 5 longer exposures.  If deep exposures were taken on nights
judged to be non-photometric, the field was repeated with short exposures
on a photometric night.  We did not reduce some of our deepest data once
we judged that what had been reduced was adequate for this study.

Most of the absolute calibration was based on repeated imaging of two fields
in Baade's Window that contained stars with single channel photometry
from \cite{fw87}.  One of these ``standard'' fields contains stars B20, 28, 31,
36, and 38; the second contains stars B143, 145, 158, 159, 162, 163, and 169.
These fields provided calibration for the nights 920714, 920715, and 920718.
On 930706 a frame was taken of the globular cluster M5 which contains two 
stars measured by \cite{fro83}.  For 930707 only a single frame of one of
Baade's Window fields was available for calibration.  On these same
nights standards from \cite{eli82} were also observed, but since our 
main objectives for this preliminary analysis involved relative photometry,
we judged the technique just described to be adequate.  Stellar photometry in
each field was carried out with DoPHOT (\cite{sch93}), while the calibration
transfers were done with the IRAF task qphot.  The photometric uncertainties
based on scatter in the repeated observations of stars in the standard fields
are about 0.04 or 0.05 magnitudes.

\section{Analysis and Discussion}

\subsection{Reddening}

We determined the average extinction and reddening to each field by assuming
that the upper giant branch in each field is similar to the upper giant
branch of Baade's Window (Tiede {\it et al.} 1995).  For Baade's Window,
the upper giant branch is very nearly linear and is defined by those
stars with $8.0 \leq K_0 \leq 12.5$.  The bright cut-off is set to
exclude luminous asymptotic giant branch (AGB) stars, while the faint
cut-off excludes clump stars.  In this magnitude range an analytic
expression for the upper giant branch is
\begin{equation}
-0.113(\pm 0.005) K_0 + 2.001(\pm 0.052) = (J-K)_0
\end{equation}
where the coefficients and uncertainties were derived by error--weighted
least--squares fitting of a line to the data from the Tiede
{\it et al.} dereddened photometry for Baade's Window.  In addition to this
equation for the upper giant branch, we used the relation between extinction
and reddening found by \cite{mat90}
\begin{equation}
A_K = 0.618 E(J-K)
\end{equation}
to establish the reddening vector in each field.

With Eqns. 1 and 2, we derived an estimate of the average reddening in each 
subfield by calculating the shift required in $K$ and $J-K$ along the reddening
vector, to force each star to fall on the Baade's Window giant branch.
Since many of the fields have foreground stars which are located in front of 
at least some of the reddening, we used an iterative process to exclude
these stars from the reddening estimate.  The mean and standard deviation 
were first calculated with all stars in a field.  The data was 
$2\sigma$-clipped and the mean and deviation recalculated.  Additional 
iterations did not exclude any additional stars except in the most reddened 
fields (c2 and c3) which require 4 iterations.  This procedure could tend 
to underestimate the true reddening dispersion in the fields.  The
final reddening for each subfield (columns 2 and 3 of Table~\ref{reddat})
was calculated from the mean of these individual shifts.  As can be seen in
Table~\ref{reddat}, the reddenings derived for each subfield within a given
field agree very well.  The uncertainties in columns 2 and 3 are based on
the star to star scatter in each of the subfields.  The averages of the
subfield values of $A_K$ and the dispersion in these averages are given
in columns 4 and 5, respectively.  In every case these dispersions are
smaller than the error in the mean values (col. 6) based on the star to
star scatter.

Finally, in addition to the upper and lower magnitude cuts and the
sigma-clipping, in the most reddened fields the $J$ exposures did not go 
deep enough to sample the entire upper giant branch.  This selection 
effect causes redder stars to be systematically excluded from the sample for
such fields and would result in an artificially low reddening estimate.
For these fields we raised the lower magnitude limit up to a level
for which there was no selection effect along the upper giant branch.
The last column of Table~\ref{reddat} lists the lower $K$-magnitude limit used
for these fields.  Since the upper giant branch is linear in the $K$, $J-K$ 
plane, this change in the faint magnitude limit
for the upper giant branch does not systematically effect either the reddening
estimate or the giant branch slope estimate below (see section 4.2 of 
Tiede {\it et al.} 1995 for a discussion).

Representative dereddened CMDs are shown in Figures~\ref{c1}--\ref{g028}.
Figure~\ref{c1} shows CMDs of subfields c1a3 (left panel) and c1b (right
panel).  The c1 field is at a low latitude and has moderate reddening.
The figure
illustrates various aspects of our reddening determination procedure.
The solid line in both panels is the dereddened upper giant branch from 
Baade's Window.  The c1 stars have been dereddened by the amount indicated in 
Table~\ref{reddat}.  Stars fainter than the dotted line and brighter than
the top of the solid line, were excluded for the reasons given above.
Stars falling more than $2\sigma$ from the line
in color were excluded primarily to select against foreground stars.  In
the right panel, stars with $K_0 > 11.07$ (dotted line) were excluded
because the $J$
frame was not deep enough to give a complete sample unbiased in color.
The remaining stars, indicated by filled points, were the ones used to estimate
the average reddening in the field.  As can be seen in this moderately 
reddened field, these latter stars, dereddened by the average reddening,
fall reasonably well along the upper giant branch defined by the Baade's
Window giants.  This morphological agreement with the Baade's Window
giant branch, coupled with the excellent agreements of reddening estimates
of different subfields in the same field gives confidence that our method
is providing a reasonable estimate of the average reddening.

Figure~\ref{c2} shows CMDs of subfields c2a (left panel) and c2b (right
panel).  The c2 field is our most heavily reddened.  All of the same
exclusions of stars where made as in the previous example, but note
that in this field the lower magnitude limit necessary, due to incompleteness
in the $J$ exposure, is much more extreme, ($K_0 > 9.49$ for c2a and
$K_0 > 9.41$ for c2b, see dotted lines in Figure~\ref{c2}).  However, even in 
this worse case, the extinction estimates in the two subfields differ only
by 0.08 magnitudes in $K$ (see Table~\ref{reddat}).  This is only half
of the error in the average value.  Note that due to the heavy reddening in
the c2 field, Figure~\ref{c2} extends to much brighter magnitudes than either
Figure~\ref{c1} or Figure~\ref{g028}.  This is caused by the extinction
effectively lowering the saturation to brighter $K_0$ magnitudes.  The
number of stars brighter than the tip of the first ascent giant branch
is discussed in section 3.6 below.

In contrast, Figure~\ref{g028} shows CMDs of subfields g0-2.8a (upper left),
g0-2.8b (upper right), g0-2.8c (lower left) and g0-2.8d (lower right).  Field
g0-2.8 has the deepest photometry (b and d subfields) and is
typical of all the g field CMDs.  Due to the drop-off in stellar 
density moving away from the Galactic center, the upper giant branches
are not as well populated as in the c fields; however, the completeness
in the $J$ exposures at the lower end of the upper giant branch
more than makes up for the stars lost due to the density fall-off.
Additionally, the decrease in differential reddening in these higher
latitude fields, allows a more precise determination of the average
reddening.

\subsection{Differential Reddening}

Near the Galactic center, differential reddening due to non-uniform
distribution of the intervening dust on small spatial scales makes the study
of the stellar population difficult even in the infrared
(e.g., Narayanan {\it et al.} 1996; \cite{dav98}).
Figure~\ref{Akscat} is a plot of $\langle A_K \rangle$ for each subfield
versus the 1-$\sigma$ scatter about the mean (col. 2 of Table~\ref{reddat})
of the individual stars.  Since the relative
star to star photometric uncertainty is similar for all fields, the 
difference in scatter from one
field to the next is an indicator of true differential reddening
in each field.  From this plot, it appears that the amount of differential
reddening in a field is directly proportional to the average reddening:
the more reddening present in a field, the more patchy that
reddening is.  A least-squares fit to the points in Fig.~\ref{Akscat}
(shown by the straight line) is given by;
\begin{equation}
\sigma(A_K) = 0.056(\pm0.005) \langle A_K \rangle + 0.043(\pm0.005)
\end{equation}
where the uncertainties in the coefficients are the formal uncertainties
in the fit.  The projected scatter at zero extinction, 0.043, is comparable
to the typical uncertainty in the photometry, about 0.05 magnitudes.

We searched for a dependence of the difference in reddening between two
stars and their spatial separation but found none.  This indicates that
the typical scale length for significant changes in extinction due to
the intevening clouds is comparable to or smaller than a few arc
seconds.

\subsection{Field Metallicities}

\cite{kuc95} determined that the slope of the upper giant branch in
metal-rich globular clusters is closely correlated with the metallicity
of the clusters.  \cite{tie95} demonstrated that when applied to
Baade's Window giants, this relationship between giant branch slope and
metallicity yielded a value for metallicity in agreement with the mean
[Fe/H] value determined by \cite{mcw94} for K giants in  Baade's
Window.  \cite{tie97} recalibrated the relationship for bulge stars and
extended it to open clusters.  Through this series of papers, the
method has been found to be robust to errors in reddening and to be
supported by theoretical considerations.  We note though that
uncertainties in the color transformations between the different
detector systems used in the different studies could affect the slope
comparisons for fields as heavily reddened as those discussed here.
This is because the change in $J-K$ color over the part of the giant
branch under consideration is about 0.5 magnitudes.

In what follows we will refer to the metallicities we determine as
[Fe/H] values although, as pointed out in the Introduction, differences
in elemental ratios with respect to iron between bulge fields and
globular clusters could affect the values we
derive.  We derive an estimate of the mean [Fe/H] for each field by
allowing the slope of the dereddened giant branch to be a free
parameter and then use least-squares to fit a line to the stars selected as
upper giant branch stars in section 3.1.  Equivalently, we could have
determined the slope of the giant branch from the observed colors since
{\it the slope is independent of reddening}, only the intercept will
change.  However, the reddening calculations described in section 3.1
also indicated likely foreground stars for elimination from the
calculations and facilitated choosing a lower limit which excluded
clump stars.

The derived $\langle$[Fe/H]$\rangle$ estimates for each field, the
individual mean [Fe/H] values for each subfield, along with the slopes,
intercepts and number of selected stars for each subfield are tabulated
in Table~\ref{metdat}.  Some subfields were observed more than once, on
different runs or different nights (see Table~\ref{obsdat} and
Table~\ref{reddat}).  Only the best exposures were used to calculate
mean metallicities.  Column 1 of Table~\ref{metdat} indicates these
exposures.  Fields c2 and c3 were excluded from this analysis because
the very large extinction in these fields -- both absolute and
differential -- resulted in a strongly truncated observed giant branch
which in turn made both slope determination and foreground-star
exclusion excessively uncertain.  Fields g4-1.3c,d were excluded
because the small total number of stars made it difficult to reliably
evaluate foreground contamination.

To calculate [Fe/H] in each subfield we used the relation for bulge stars
from \cite{tie97}: 
\begin{equation}
{\rm [Fe/H]} = -1.692(\pm 0.500) -13.613(\pm 5.118)\times{\rm(GB~slope)}.
\end{equation}
This recalibration of the relation derived by \cite{kuc95} is
based on only 5 minor-axis bulge fields.  However, as demonstrated in 
Figure 2 of \cite{tie97}, a recalibration of the original metal-rich
globular cluster relation is necessary for bulge stars, especially
for [Fe/H] $\lesssim -0.4$.

Columns 5--8 of Table~\ref{metdat} present the results of these
calculations.  Column 5 lists the [Fe/H] estimates for each subfield
calculated from the above relation.  Column 6 is the error-weighted
average metallicity for each field calculated from the metallicities of
each associated subfield.  Column 7 is the standard deviation in the
subfield metallicity values and column 8 is the formal error propagated
through the calibration equations.  Most of the standard deviations are
not very meaningful since they are each based on only 2 subfields;
however, so as not to under estimate the errors in the metallicity
values, for subsequent analysis in cases where the standard deviation
is larger than the formal error, the standard deviation was used for
error analysis.

In theory we could use the newly derived slopes to iterate on our
reddening estimates.  In practice this will not work since the
difference in {\it true} (as opposed to observed) slopes for each field
is actually quite small.  If we exclude the fields with the highest
reddening and greatest uncertainties in their slopes, c2 and c3, then
between b$= -0.8^\circ$ and $-2.8^\circ$ the change in average slope is
only 0.01.  The resulting change in the projection of the reddening
vector onto the giant branch will also be small compared with the
uncertainties.  Thus the derived dispersion in the extinction will not
change significantly.  Furthermore, since we will still not know the
change in the zero point, we could not calculate what the change in
reddening would be, although it again would be expected to be small.
Although an iterative approach might be expected to work best for the
most heavily reddened fields, these also have the most poorly
determined slopes due to the relatively small stretch of the giant
branch that could be observed in these fields.

\subsection{Minor-Axis Metallicity Gradient}

\cite{tie95} published metallicity values derived from the GB
slope-metallicity method for minor-axis fields spanning $-0.2^\circ
\leq {\rm b} \leq -12.5^\circ$.  However, only one field, near the
globular cluster Liller 1, fell within 2 degrees of the Galactic
center.  With our newly derived metallicity values we can, for the
first time, determine the minor-axis metallicity gradient for the inner
bulge, $-0.8^\circ \leq {\rm b} \leq -2.8^\circ$.  We exclude the
off-axis fields, g1-1.3, g2-1.3, g3-1.3, and g4-1.3, from consideration
in calculating this gradient.

Figure~\ref{innergrad} is a plot of derived $\langle$[Fe/H]$\rangle$
versus galactic latitude, $b$, for the inner bulge fields that are
close to or on the minor axis.  The line is an error-weighted
least-squares fit to the points.  The slope of the line is $-0.085 \pm
0.033$ dex/degree with an intercept of $-0.019 \pm 0.066$ dex.  This
metallicity gradient agrees with the value found by \cite{tie95},
$-0.060 \pm 0.033$, based primarily on fields at greater Galactocentric
distance.

Since there is good agreement between the present determination of the
minor-axis metallicity gradient for the inner bulge and that of
\cite{tie95} for the outer bulge, we combine the two determinations to
produce one set of metallicity estimates spanning $-0.2^\circ \leq {\rm
b} \leq -10.25^\circ$. The values from \cite{tie95} are recalibrated to
the new slope--metallicity relation for bulge stars (Eq. 3).  The
galactic latitudes and $\langle$[Fe/H]$\rangle$ values are given in
Table~\ref{gradmet}.  We exclude the field at b$= -12.5^\circ$ because,
as shown in \cite{tie97} (see also Minniti {\it et al.} 1995),
its population is
likely dominated by halo/thick disk stars rather then bulge stars.
Figure~\ref{totgrad} plots the $\langle$[Fe/H]$\rangle$ estimates from
Table~\ref{gradmet} versus galactic latitude.  The filled circles are
data from this study while the open circles are the recalibrated data
from \cite{tie95}.  The line is the calculated error-weighted
least-squares fit to the 12 points.  This fit gives the metallicity
gradient for the entire bulge population along the $b < 0^\circ$ minor-axis
to be $-0.064 \pm 0.012$ dex/degree with an intercept (predicted [Fe/H]
value at the Galactic center) of $+0.034 \pm 0.053$ dex.  Values found
in previous studies for the {\it outer} bulge metallicity gradient are
in reasonable agreement with our new value for the entire bulge (see
Table~\ref{grads}).

As just noted, our data predict that the bulge population near the
Galactic center, presumably the most metal-rich, should have a
metallicity value close to solar.  This prediction along with the
$\langle$[Fe/H]$\rangle$ values derived along the minor axis are in
disagreement with studies such as \cite{bla84} and \cite{bla89} which
suggested super-solar metallicities based on the large number of
M~giants present in the bulge, and analyses such as those by
\cite{fw87}, \cite{ric88} and Terndrup {\it et al.} (1990,1991) which
also suggested super-solar metallicities for bulge stars in Baade's
Window based on CO index strengths and medium-resolution spectral
measurements of the TiO band.  However, this prediction is in agreement
with later studies such as the high-resolution spectroscopy of K giants
in Baade's Window by \cite{mcw94} and the conclusion by \cite{hou95}
that the stars in the Galactic bulge are more similar to the sub-solar
metallicity stars found in lower-mass elliptical galaxies than the
super-solar metallicity stars found in massive elliptical galaxies.
This evidence suggests that bulge stars are selectively enriched in
certain elements (e.g., Ti and O) while their overall metallicities, as
measured by the GB slope - metallicity relation (cf. Tiede {\it et al.} 1995),
are just below solar.

Most previous photometric and spectroscopic studies, both in the red
and near-IR, of the M giants in the Galactic bulge have not given a
hint of a spread in metallicity at any given latitude, particularly in
Baade's Window (e.g. \cite{fw87}, Terndrup {\it et al.} 1990, 1991).
In contrast, high resolution optical studies of K giants in Baade's
Window (e.g. Rich 1988) show the existence of a spread in [Fe/H] of
nearly a factor of 100.  With the new data presented here it would be
difficult to distinguish a spread in metallicity from dispersion of
reddening.  Nevertheless, the fact that the dispersion in $(J-K)$ is
about 0.10 mags or less for many of the fields we have surveyed,
suggests that a metallicity dispersion approaching a factor of 100
would not be allowed by our new data for the inner bulge.  
Based on observations of globular cluster giant branches and of fields
in the outer bulge, we estimate that the allowable range in metallicity
dispersion for a give field is likely to be no more than a factor of 10 
(\cite{fro90}). 

Measuring the $\langle$[Fe/H]$\rangle$ of the bulge population at the
Galactic center is problematic due to the extremely high reddening and
reddening dispersion as demonstrated, for example, by \cite{blu96a} and
\cite{nar96}.  Additionally, \cite{blu96b} have shown that the Galactic
center region contains stars from distinct star-formation epochs with
ages from $\sim7$ Myrs (the most recent) to $\sim10$ Gyrs (the bulge
population).  However, \cite{ram97} have begun to measure metallicities
for Galactic center stars based on high-resolution spectral analysis
and determine a $\langle$[Fe/H]$\rangle$ value (based on measurements
of actual Fe absorption lines) of $-0.07 \pm 0.11$ for 10 stars of
various ages, including 2 which are super giants and so belong to the
youngest epoch of star formation.  It appears that while the bulge may
have enriched faster than the halo or disk, even the most recent epoch
of star formation in the Galactic center did not produce an easily
detectable population of super-solar metallicity stars.

We conclude this subsection by noting the following: The medium and high resolution optical spectroscopy for
Baade's Window K giants as well as similar data for higher latitude
fields probably give the most reliable indicators of stellar abundances in the
outer bulge.  Therefore, there is considerable need to understand why
most of the analyses of the M giants, both optically and in the near-IR
have yielded findings at odds with the K giant studies, in particular
the mean value for [Fe/H] and its dispersion.  Resolution of these
conflicting results is especially important because of the fact that
only near-IR data can be obtained for much of the inner bulge.

\subsection{Major-Axis Metallicity Gradient}

With the increasing complexity of kinematic models of the inner galaxy and
bulge, it is not obvious whether a major-axis metallicity gradient is
expected or not.  With the asymmetric kinematics suggested by models
such as \cite{zha96}, line-of-sight distributions of stars may peak in
density at different Galactocentric radial distances complicating the
projected composite population seen at particular $(\ell, b)$. 
To investigate this matter, we plot $\langle$[Fe/H]$\rangle$ versus
galactic longitude of our 5 fields with $b = -1.3$, in Figure~\ref{major}.
The line is an error-weighted least-squares fit to the points.
The data only span the inner $4^\circ$ of longitude in the first Galactic
quadrant, however, we find no evidence for a metallicity gradient; the
slope of the line is $0.004\pm 0.080$ dex/degree. 

\subsection{Young Population in the Inner Bulge}

While many studies have now identified bright, young super giants and
luminous AGB stars in the region within a few arc minutes of the
Galactic center  (e.g., Blum {\it et al.} 1996a,b; Narayanan {\it et
al.} 1996), none have made a study of the extent of the spatial
distribution of these stars in the inner bulge.  Our inner fields; c1,
c2, and c3, have a large number of stars brighter than the giant branch
tip (for $R_0 = 8.0$ Kpc, $K_0 \sim 8.0$).  While some of these bright
stars are bulge asymptotic giant branch stars or foreground giants, the
large number suggests that many could be members of a young population.

Since our various fields have different areas and are complete to
different $K_0$, we examined the distribution of these bright stars by
calculating the ratio of the number of stars with $K_0 < 8.0$,
including all saturated stars, to the number of old bulge population
upper giant branch stars with $K_0$ magnitudes in the range $8.0 \leq
K_0 \leq 9.4$.  The $9.4$ magnitude faint limit was selected so that
all of our fields would be complete in this range.  The calculated
ratios along with the angular radial distance from the Galactic center
are tabulated in Table~\ref{satdat} and displayed in Figure~\ref{sat}.

Table~\ref{satdat} lists each field, Galactic coordinates of the center
of each field, the angular radial distance from the Galactic center,
the number of stars in each $K_0$ magnitude range, and the ratio of the
number of bright stars to the number of faint stars.  The error
associated with each ratio is the formal counting error propagated
through the ratio.  Figure~\ref{sat} is a plot of these ratios versus
angular distance from the Galactic center.  The solid points are the
minor-axis fields.  The open squares are the major-axis fields.  The
solid square is field g0-1.3 where the major- and minor-axis field
groups intersect.  The $\otimes$ is the ratio for Baade's Window from
the luminosity function presented in \cite{tie95}.

While the uncertainties in individual ratios are generally quite large,
we note two trends.  Along the minor axis (solid points) the ratio
decreases from $\sim0.6$ for the inner most fields to $\sim0.3$ for the
g0-1.3 field, $1.3^\circ$ from the Galactic center.  At radial
distances greater than $1.3^\circ$, the ratio remains generally
constant around $0.3$ for both the minor-axis fields and the fields
which parallel the major axis.  In fact, the average value for all of
these fields is about $0.3$. The horizontal line in Figure~\ref{sat} is
drawn at this value.  This is also the value found in Baade's Window
(the circled ``x'' in Fig.~\ref{sat}) which is not known to contain any
luminous,  young stars (DePoy {\it et al.} 1993; Blum {\it et al.}
1996a). Therefore, it seems likely that the young population is
confined to within $\sim1^\circ$ of the Galactic center and bright
stars outside this region are old bulge population asymptotic giant
branch stars.

Finally, we note that the presence of a significant number of
relatively young AGB stars in the two innermost ``c'' fields (c2 and
c3) could contribute to the scatter in the CMDs for these fields.
However, based on observations of clusters with a wide range in age in
the Large Magellanic Cloud (Frogel, Mould, \& Blanco 1990), the
expected scatter in $(J-K)_0$ due to a wide age spread is about 0.1
mags, less than half of that observed in c2 and c3 (Fig. 3 and Table
2).

\section{Conclusions}

In this work we analyze near-IR, $JK$, photometry  for 11 fields in the
inner $4^\circ$ of the Galactic bulge.  Seven of these fields are along
the minor axis spanning from $b = -2.8^\circ$ to $-0.2^\circ$.  The other
fields are arranged parallel to the major axis at $b=-1.3^\circ$.

We determined the average extinction and reddening in each field by
assuming the old bulge population present in each window has an upper
giant branch similar to that found in $K, J-K$ color magnitude diagrams
of Baade's Window.  We constructed $K, J-K$ CMDs for each field and
calculated the average shift along the reddening vector required for
the giant branch to coincide with the Baade's Window giant branch.  The
resulting extinctions, displayed in Table~\ref{reddat}, ranged from
$E(J-K) = 0.27 \pm 0.05$ to $2.15 \pm 0.16$.

We found that the dispersion in $K$ and $J-K$ around each giant branch
is larger than can be accounted for by photometric uncertainty and that
the size of this dispersion is directly proportional to the average
extinction in the field.  Thus, differential reddening seems to
directly correlate with increased extinction in the fields.

Using the reddening estimates to exclude likely foreground and clump
stars, we derived an estimate of the $\langle$[Fe/H]$\rangle$ for each
field using the correlation between the slope of the upper giant branch
and [Fe/H] derived by \cite{kuc95} and recalibrated for bulge stars by
\cite{tie97}.  With the exception of the fields with large differential
reddening for which the method is unreliable, we found
$\langle$[Fe/H]$\rangle$ to generally be just below solar,
$\langle$[Fe/H]$\rangle \sim -0.2$.

We derived the gradient in mean metallicity along the minor axis of the
inner Galactic bulge in the range $-0.2^\circ \geq b \geq -2.8^\circ$.
This result was combined with that of Tiede {\it et al.} (1995) based
almost completely on data for fields exterior to the present ones,
extending to $b=-10.25^\circ$, and derived a metallicity gradient for
the entire bulge minor axis.  This gradient is $-0.064 \pm 0.012$
dex/degree with an intercept (predicted value of [Fe/H] at the Galactic
center) of $+0.034 \pm 0.053$ dex.  This near-solar value for the
Galactic center agrees with the recently determined value for
$\langle$[Fe/H]$\rangle$ for the bright stars in the Galactic center
based on high resolution near-IR spectroscopy.  We also examined  5
fields distributed in galactic longitude but along a line of constant
latitude of $b=-1.3^\circ$, and found no evidence of a major-axis
metallicity gradient.

Finally, we observed that there were a significant number of stars
brighter than the giant branch tip (for $R_0 = 8.0$ Kpc, $K_0 \approx
8.0$) in our inner most fields, but fewer in the more distant fields.
To examine the distribution of these bright stars, we calculated the
ratio of the number of stars with $K_0 < 8.0$ to the number of stars
with $8.0 \leq K_0 \leq 9.4$.  We found that the ratio quickly drops
from $\sim0.6$ in the inner most fields to $\sim 0.3$ in the field
$1.3^\circ$ from the Galactic center.  This $0.3$ value is generally
characteristic for all fields further from the Galactic center,
including Baade's Window which is known to contain no bright young
stars like those found at the Galactic center.  Thus we conclude that
the young Galactic center population does not extend out beyond
$1^\circ$ from the Galactic center.  Beyond this distance, stars
brighter than the giant branch tip are likely asymptotic giant branch
stars associated with the old bulge population.

\acknowledgements

We thank Kris Sellgren,  Solange Ram\'{\i}rez, and Don Terndrup for
helpful comments and discussion.  JAF thanks Leonard Searle for the
observing opportunities provided by a Visiting Research Associateship
at LCO where the data were obtained.  This research was supported in
part by NSF grant AST92-18281 to JAF.  We are particularly grateful to
S. E. Persson for the use of IRCAM (built with NSF funds) at LCO and
also thank Miguel Roth, Bill Kunkel, and members of the LCO support
crew for their able assistance in the set up and operation of IRCAM and
the duPont telescope.  We appreciate the advice given by Paul Schechter
on the use of DoPHOT which was developed with the aid of NSF grant
AST83-18504.

\clearpage

\begin{figure}
\figurenum{1}
\caption{Galactic positions of our 11 fields.  The dotted lines
represent the major and minor Galactic Axes.  The location of Baade's
Window is shown for reference.}
\label{fields}
\end{figure}

\begin{figure}
\figurenum{2}
\caption{$K_0, (J-K)_0$ color magnitude diagrams of two of the subfields from
our c1 field; this is our lowest latitude field on the minor axis.  It has
moderate reddening, $E(J-K) = 1.67 \pm 0.15$.  The solid line is the 
dereddened giant branch from Baade's Window, used to estimate reddening.  The
dotted horizontal line is the completeness limit in the $J$ frame.  The
open points were excluded from the reddening determination (see text for
details).}
\label{c1}
\end{figure}

\begin{figure}
\figurenum{3}
\caption{$K_0, (J-K)_0$ color magnitude diagrams of two of the subfields from
our c2 field.  Field c2 is the most heavily reddened field in our data and
contains a large number of stars brighter than the giant branch tip.
Lines and points are the same as in Figure~\ref{c1}.}
\label{c2}
\end{figure}

\begin{figure}
\figurenum{4}
\caption{$K_0, (J-K)_0$ color magnitude diagrams of the four subfields from
our g0-2.8 field.  g0-2.8 is typical of all the g field CMDs.  The b and d
subfields have deeper photometry than the a and c subfields.  The small
number of stars along the upper giant branch relative to the c fields is 
due to decreased stellar density at larger galactic latitudes.  Lines and
points are the same as in  Figure~\ref{c1}.}
\label{g028}
\end{figure}

\begin{figure}
\figurenum{5}
\caption{Mean extinction at $K$ versus the 1-$\sigma$ scatter about the
mean for our subfields.  Note that the scatter in $A_K$ is directly
proportional to $A_K$ in these fields.  The line is a least-squares fit
to the data.}
\label{Akscat}
\end{figure}

\begin{figure}
\figurenum{6}
\caption{Mean field metallicity versus Galactic latitude in degrees along the
minor-axis for our minor-axis fields.  Data are from Table~\ref{gradmet}.
The line is an error-weighted least-squares fit to the points.  The slope of
the line is $-0.085 \pm 0.033$ dex/degree.}
\label{innergrad}
\end{figure}

\begin{figure}
\figurenum{7} 
\caption{Mean field metallicity versus Galactic latitude in degrees along the
minor-axis for our minor-axis fields (open circles) and fields from 
Tiede {\it et al.} (1995) (solid circles).  The [Fe/H] values from 
Tiede {\it et al.} (1995) have been recalibrated to the 
Tiede {\it et al.} (1997) system (see 
Table~\ref{gradmet}) to be consistent with the values for our fields.  The
line is an error-weighted least-squares fit to all of the data.  The slope
of the line is $-0.064 \pm 0.012$ dex/degree with an intercept of
$+0.034 \pm 0.053$ dex.}
\label{totgrad}
\end{figure}

\begin{figure}
\figurenum{8}
\caption{Mean field metallicity versus Galactic longitude in degrees for our
fields distributed parallel to the major axis at $b = -1.3^\circ$.  The line
is an error-weighted least-squares fit to the data.  The slope of the
line is $0.004 \pm 0.080$ dex/degree, consistent with no metallicity
gradient in Galactic longitude.}
\label{major}
\end{figure}

\begin{figure}
\figurenum{9}
\caption{The ratio of the number of stars with $K_0 < 8.0$ 
to the number of stars with $8.0 \leq K_0 \leq 9.4$ versus
Galactocentric distance in degrees.  Solid points are the minor-axis
fields.  The open squares are the major-axis fields.  The solid square is 
field g0-1.3 where the major- and minor-axis field groups intersect.  
The $\otimes$
is the ratio for Baade's Window calculated from the luminosity function
presented in Tiede {\it et al.} (1995). The average for all fields with
Galactocentric distances greater than 1 degree is nearly equal to the 
Baade's Window value and is represented by the horizontal line in the figure.}

\label{sat}
\end{figure}

\begin{table}
\tablenum{1}
\label{obsdat}
\end{table}
 
\begin{table}
\tablenum{2}
\label{reddat}
\end{table}
 
\begin{table}
\tablenum{3}
\label{metdat}
\end{table}

\begin{table}
\tablenum{4}
\label{gradmet}
\end{table}

\begin{table}
\tablenum{5}
\label{grads}
\end{table}

\begin{table}
\tablenum{6}
\label{satdat}
\end{table}

\end{document}